\documentclass[intlimits,twoside,a4paper]{article}

\usepackage{graphicx}
\usepackage[cp1251]{inputenc}

\usepackage{cmpj3}
\issue{2017}{20}{4}{43602}
\doinumber{10.5488/CMP.20.43602}

\title[Composition dependence of the properties of water-DMF mixtures]
{On the composition dependence of the microscopic structure, thermodynamic, dynamic 
and dielectric properties of 
water-dimethyl formamide model mixtures. Molecular dynamics simulation results}

\author[H. Dominguez, O. Pizio]{H. Dominguez, O. Pizio\footnote{On sabbatical leave from Instituto de Quimica 
de la UNAM, corresponding author: oapizio@gmail.com.}}
\address{
Instituto de Investigaciones en Materiales, Universidad Nacional Aut\'{o}noma de M\'{e}xico,
Circuito Exterior, 04510, M\'{e}xico, D.F., M\'{e}xico
}

\date{Received August 10, 2017, in final form September 11, 2017}

\begin{document}

\maketitle

\begin{abstract}
Isothermal-isobaric  molecular dynamics simulations
have been performed to examine an ample set of properties
of the model  water-N,N-dimethylformamide (DMF) mixture as a function of composition.
The SPC-E and TIP4P-Ew water models together with two united atom models for DMF
[Chalaris~M., Samios~J., J. Chem. Phys., 2000, \textbf{112}, 8581;
Cordeiro J., Int. J. Quantum Chem., 1997, \textbf{65}, 709]  
were used.
Our principal analyses concern the behaviour of  structural
properties in terms of  radial distribution functions, and the
number of hydrogen bonds between molecules of different species as well as 
thermodynamic properties. Namely, we explore the
density,  excess mixing molar volume and enthalpy, the heat capacity 
and excess mixing heat capacity.  Finally, the self-diffusion
coefficients of species and the dielectric constant of the system are discussed.
In addition, surface tension of water-DMF mixtures has been calculated and analyzed.


\pacs 61.20.-p, 61.20-Gy, 61.20.Ja, 65.20.Jk

\end{abstract}

\section{Introduction}

N,N-dimethylformamide (DMF) is a multipurpose organic solvent with widespread applications
in chemistry and related areas of much practical importance, 
see e.g.,~\cite{kroschwitz,fiorito,ohara}.
In several processes it is used as a co-solvent. On the other hand,
as one example of amides, it is frequently chosen as a model system for peptides.
Given the importance of  DMF and of its mixtures with water and several organic solvents, 
investigation
of the properties of such complex molecular systems has been undertaken using a variety
of experimental techniques. Namely, the studies of microscopic structure have been performed 
using  neutron and electron diffraction,  NMR and dielectric spectroscopy;  data resulting 
from measurements of various properties, like density, excess molar volume and enthalpy, 
excess molar heat capacity, self-diffusion coefficients, dielectric constant have been reported in 
\cite{ohtaki,okada,biliskov,schoester,guarino,volpe,bernal,visser,cilense,kumbharkhane,checoni,gofurov,ueno,scharlin,han,garcia,shokouhi,chen,jadzyn,egorov,bai1,bai2}. 
Some experimental observations were supported by quantum chemical 
calculations~\cite{schmid1,schmid2}. 

\looseness=-1 In addition, aiming at rather general and more profound insights into the behavior 
of various properties on temperature, pressure and chemical composition of water-DMF 
mixtures and at interpretation of experimental data, computer simulation 
methods have been applied to these mixtures. N,N-dimethylformamide is an interesting molecule,
hydrophobic methyl groups can disturb the water structure. Besides, ``cross'' hydrogen bonds
are expected to form between molecules belonging to two different species.
Computer modelling studies can be classified
according to the details of the force fields describing the internal structure of the DMF molecule. 
Specifically, one set of models with either five or six force centers
(in each of models the methyl group has been considered as a single site) was explored 
in~\cite{jorgensen,yashonath,cordeiro1,cordeiro2,cordeiro3,bako,chalaris1,chalaris2,chalaris3,zoranic1,zoranic2,gao}.
Another set of apparently  more sophisticated models, i.e., at all-atom level, has been designed 
and then studied in detail in several publications~\cite{biswas,vasudevan,lei,jia,razzokov}.
A very comprehensive description of the properties of an ample set of organic liquids resulting 
from simulations with various force fields has been discussed in~\cite{fischer} (see also the file that contains a vast supporting information of \cite{fischer} and the \url{http://virtualchemistry.org}  web site).

In this work we restrict our attention solely to the modelling of water-DMF mixtures and to the 
critical evaluation of several properties as a function of composition 
resulting from isobaric-isothermal computer simulations. Moreover,
our attention is restricted to the united-atom type models for DMF molecule developed by
Cordeiro and by Chalaris and Samios. The reason of studying these models, in addition to
and in spite of the available knowledge, is that we believe several even basic issues have
not been comprehensively and critically discussed. Specifically, we calculate density
and excess mixing density, excess mixing volume and enthalpy, heat capacity and excess mixing heat
capacity, self-diffusion coefficients of species of the mixture, dielectric constant and
the excess dielectric constant, average number of hydrogen bonds, all as functions of mixture
composition. The microscopic structure is briefly discussed in terms of the pair distribution 
functions. Finally, we performed calculations of the surface tension switching to the
constant volume-constant temperature canonical ensemble.
For the majority of properties calculated, we perform comparisons with the 
experimental results. Having in mind our recent investigation of the properties
of water-methanol~\cite{galicia1,galicia2}, water-DMSO~\cite{gujt1} and water-1,2-dimethoxyethane
(DME) mixtures~\cite{gujt2}, the results of the present work are discussed in a wider context
in qualitative terms,
making comparisons  of the trends observed for different organic liquids  mixed with water.

Exploration of the behavior of water-DMF mixtures by using all-atom type models
for DMF will be presented elsewhere in a separate work. 
Our longer term objective of the project, however, is in the theoretical exploration of the 
behavior of solutes of different complexity in water-DMF and other water-organic liquid
mixtures in the spirit of important and interesting experimental 
studies~\cite{takamuku1,takamuku2,takamuku3,takamuku4}. 

\section{Models and simulation details}

Our calculations have been performed in the isothermal-isobaric (NPT) ensemble
(at 1~bar, and at a temperature of 298.15~K), unless specified in the
surface tension subsection.  We used the GROMACS
package~\cite{gromacs} version 4.6.5.
For water in the present study, the SPC-E model~\cite{spce}
and the TIP4P-Ew model~\cite{horn} were used. 
As for the DMF molecules, we explore two united-atom type models with
six force sites, H, O, C, N, each methyl group, CH$_3$ is considered as
a single site, C3. The nomenclature of the DMF models is as follows:
we use the model and notation CS2 as proposed by Chalaris and Samios,
see table~I of \cite{chalaris1}. On the other hand, the model
denominated as Cordeiro is the one proposed by this author and described in detail
in table~I of \cite{cordeiro1}. All the parameters for the intramolecular geometry 
of the rigid CS2 model are given in table II~of  \cite{chalaris1}. The model of
Cordeiro is rigid as well. For the sake of numerical convenience, in the present work 
we relaxed rigidity of bonds and angles, taking the force constants from the OPLS
data basis as appropriate. Moreover, to keep the DMF molecule planar, 
the improper dihedral angle for H-C-N-C3 and for O-C-N-C3 
has been assumed from the OPLS data basis as well.    

The geometric combination rule (rule~3 in GROMACS
nomenclature) has been applied for cross interactions in all our calculations.
The nonbonded interactions were cut off at 1.4~nm,
 and the long-range electrostatic interactions were handled by the
particle mesh Ewald method implemented in the GROMACS software package (fourth
order, Fourier spacing equal to 0.12).
The van der Waals tail correction terms to the energy and pressure were taken into account.
In order to maintain the geometry of water and DMF molecules,  LINCS
algorithm was used.

For each system a periodic cubic simulation box was set up.
The GROMACS genbox tool was employed to randomly place all particles into the
simulation box. The total number of molecules  was kept fixed at 3000.
The composition of the mixture is described by the mole fraction of DMF molecules,
$X_{\text{dmf}}$, $X_{\text{dmf}}=N_{\text{dmf}}/(N_{\text{dmf}}+N_{\text w})$.

To remove possible overlaps of particles introduced by the procedure of
preparation of the initial configuration, each system underwent energy
minimization using the steepest descent algorithm implemented in the GROMACS
package. Minimization was followed by a 50~ps NPT equilibration run at 298.15~K and 1~bar using a timestep of 0.25~fs.
We applied the Berendsen thermostat and  barostat with $\tau_T = 1$~ps and $\tau_P = 1$~ps
during equilibration. Constant value of $4.5 \times 10^{-5}$~bar$^{-1}$ for the 
compressibility of the mixtures was set up.
The V-rescale thermostat and Parrinello-Rahman
barostat with $\tau_T = 0.5$~ps and $\tau_P = 2.0$~ps
and the time step 2~fs were used during all production runs. 
To test the thermostat and barostat, we have obtained 88.5~J/mol$\cdot$K for the heat
capacity of the SPC/E model, this value being close to 86.6~J/mol$\cdot$K reported by Vega et
al.~\cite{vega} (experimental result is 75.3~J/mol$\cdot$K). 
On the other hand, we obtained 155.55~J/mol$\cdot$K for
the heat capacity of the DMF for the CS2 model, 
and 158.78~J/mol$\cdot$K for the Cordeiro model,
the experimental values reported from independent measurements are as follows:
150.16~J/mol$\cdot$K \cite{checoni}, 150.8~J/mol$\cdot$K \cite{visser}. 
Moreover, the DMF molar volume coming out from our calculations is
76.53~cm$^3$/mol (CS2  model)
and 78.23~cm$^3$/mol (Cordeiro model), the deviation from the experimental 
result, 77~cm$^3$/mol, is very small in the case of CS2.
Berendsen type control of temperature and pressure is not satisfactory in this aspect.

Statistics for each mole fraction for some of the properties were collected over
several 10~ns NPT runs, each started from the last configuration of the
preceding run. The total trajectory was not shorter than 60~ns. Actually, the heat capacity
and the dielectric constant are the most time demanding properties.

\section{Results and discussion}

\subsection{Density, mixing properties}

\begin{figure}[!b]
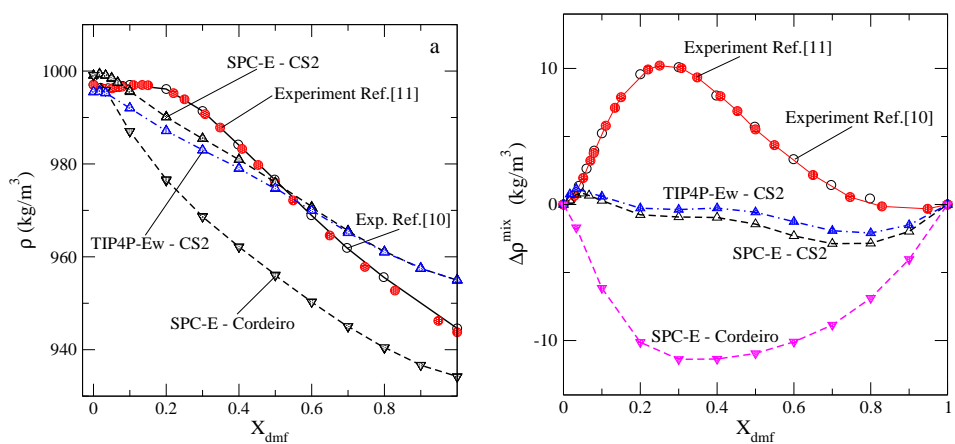

\begin{center}
\includegraphics[width=6cm,clip]{fig1a.eps} \quad
\includegraphics[width=6cm,clip]{fig1b.eps}
\end{center}
\caption{\label{fig1}(Color online) Panel (a): Composition dependence of the density of
water-DMF mixtures
from constant pressure-constant temperature simulations ($T = 298.15$~K, $P = 1$~bar)
in comparison with the experimental data from~\cite{bernal,visser}.
Panel (b): Excess mixing density on composition.}
\protect
\end{figure}

We begin a discussion of the results by presenting the dependence of density of water-DMF
mixture with united-atom type models for DMF on composition, panel~(a) of figure~\ref{fig1}. In spite 
of a meticulous search, we have not found previous comparisons of this sort. Two water models and two
DMF models are involved. The experimental data are taken from \cite{bernal,visser}.  
Experimental results show a peculiar behavior of mixture density at low values of $X_{\text{dmf}}$,
namely while DMF is added to water, the mixture density exhibits a weakly pronounced maximum,
at $X_{\text{dmf}} < 0.2$, the density preserves values close to the density of pure water,
if $X_{\text{dmf}}$ is low. Actually, $\rho(X_{\text{dmf}})$  decreases with DMF concentration 
only for  $X_{\text{dmf}} > 0.2$. A combination of SPC-E and Cordeiro model leads to a pronounced
underestimation of density in the entire composition range. On the other hand,
the SPC-E-CS2 as well as TIP4P-Ew-CS2 exhibit a more satisfactory behavior. 
Both models underestimate the values of density of the mixture at small
values of $X_{\text{dmf}}$ and slightly overestimate the density for high values of $X_{\text{dmf}}$.

It is common to characterize the behavior of various properties of binary mixtures in
terms of excess mixing contribution as follows,
\begin{equation}\label{eq1}
 \Delta Y^{\text{mix}} = Y_{\text m}- [X_{1}Y_{1}+(1-X_{1})Y_{2}],
\end{equation}
where $Y_{\text m}$ refers to the mixture at a given composition, $Y_i$ ($i=1,2$) are the
values for pure components, $Y$ can be any property of the mixture,  
e.g., density or volume or enthalpy or another one.

Here, to begin with, we obtain
a finer insight into the decay of density from pure water to pure DMF. It is provided 
by the results  given in panel~(b) of figure~\ref{fig1}. Namely, the deviation of mixture density from an ideal
type behavior is closer to the experimental trends if the SPC-E or TIP4P-Ew water models
are combined with the CS2 model for DMF. On the other hand, predictions of the SPC-E
combined with Cordeiro model are not satisfactory. Positive deviation of excess mixing
density is similar to what was observed in mixtures of alcohols with water in the
work of Wensink~\cite{wensink}. In fact, the behavior of excess mixing density can indicate
the accuracy of theoretical curves, but it does not provide explicit
insights into the mixing trends; methanol or ethanol mix with water perfectly well,
but the $\Delta \rho$ is positive in the entire range of composition.

\begin{figure}[!t]
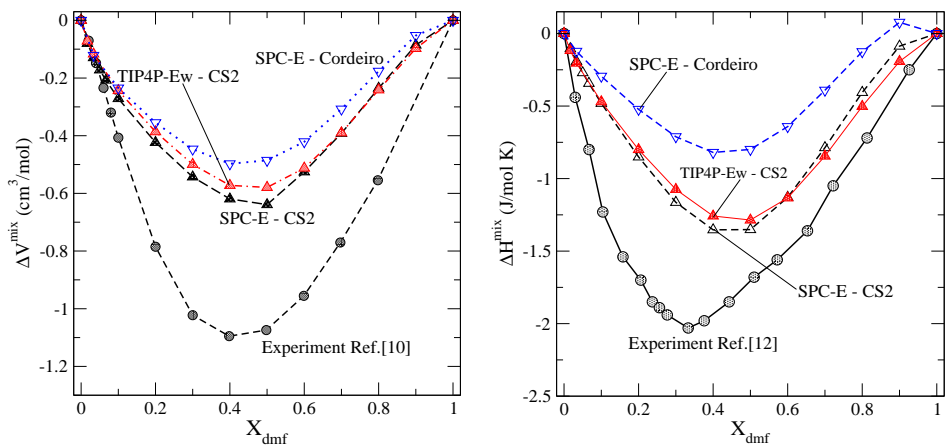

\begin{center}
\includegraphics[width=6cm,clip]{fig2a.eps}\quad
\includegraphics[width=6cm,clip]{fig2b.eps}
\end{center}
\caption{\label{fig2}(Color online) Excess mixing molar volume and excess mixing enthalpy of water-DMF mixtures
on composition.}
\protect
\end{figure}

The excess mixing volume and enthalpy are given in two panels of figure~\ref{fig2}. From panel~(a)
of this figure we learn that the excess mixing volume describes a contraction
of the mixture upon composition qualitatively well, although the magnitude of $\Delta V^{\text{mix}}$ is
underestimated. The minimum value from the simulations is at $X_{\text{dmf}} \approx 0.5$,
whereas the experimental result attains minimum at $X_{\text{dmf}} = 0.4$.
The closest, to the experiment, set of theoretical data comes out from the SPC-E-CS2 model. 

One can see quite similar trends  in the behavior of excess mixing enthalpy, panel~(b) of figure~\ref{fig2}.
The values for $\Delta H^{\text{mix}}(X_{\text{dmf}})$ are underestimated by all the models considered and
the minimum occurs at a slightly higher $X_{\text{dmf}}$ (between 0.4 and 0.5), in comparison to 
the experimental result at $X_{\text{dmf}} \approx  0.35$.
Again, the SPC-E-CS2 model is better comparing to, e.g., SPC-E-Cordeiro version of the model.
In general, one can conclude that the united-atom type models of this work underestimate
geometric and energetic trends for mixing of species given by $\Delta V^{\text{mix}}$ and by 
$\Delta H^{\text{mix}}$ in the entire interval of mixture compositions. The agreement with the
experimental results is qualitative for both properties in question.

\begin{figure}[!t]
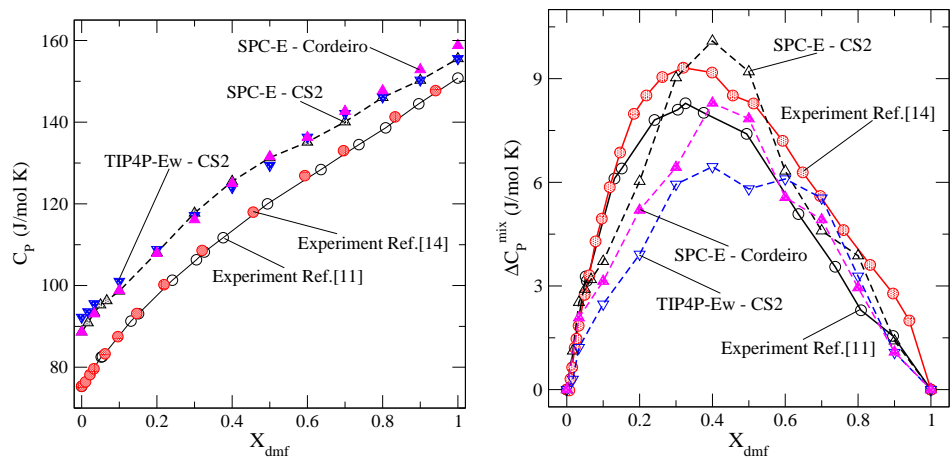

\begin{center}
\includegraphics[width=6cm,clip]{fig3a.eps}\quad
\includegraphics[width=6cm,clip]{fig3b.eps}
\end{center}
\caption{\label{fig3}(Color online) Panel~(a): Composition dependence of the heat capacity at constant pressure of
water-DMF mixtures
from constant pressure-constant temperature simulations ($T = 298.15$~K, $P = 1$~bar)
in comparison with the experimental data from~\cite{checoni,visser}.
Panel~(b): Excess mixing constant pressure heat capacity on composition in comparison with
experimental data.}
\end{figure}

In addition, it is worth studying the energetic trends of mixing of species in terms of 
fluctuations. Namely, in figure~\ref{fig3} we explore the dependence of the constant pressure heat capacity 
and the corresponding excess property as functions of mixture composition. Actually,
the situation is rather satisfactory. From panel (a) of figure~\ref{fig3}, we learn that trends of
dependence of $C_P$ on $X_{\text{dmf}}$ are well reproduced by all the models studied, just the
theoretical curves are shifted upward in comparison to the experimental dependence.
The shape and magnitude of $\Delta C_P^{\text{mix}}$ is reproduced by the simulated models very well.
Moreover, the maximum is predicted at 0.4 whereas in the experiment it is at $\approx 0.3$.
Seemingly, the SPC-E-CS2 combination of models is the best in this aspect.

After exploring thermodynamic aspects of mixing, we proceed to the brief description of
the microscopic structure in terms of the pair distribution functions and hydrogen bonding
between molecules.

\subsection{Pair distribution functions, hydrogen bonding} 

All the pair distribution functions discussed below result from
simulations of the SPC-E-CS2 model. We have checked, however, that other models, like
TIP4P-Ew-CS2 and SPC-E-Cordeiro, yield a qualitatively similar picture of the trends observed.

\begin{figure}[!b]
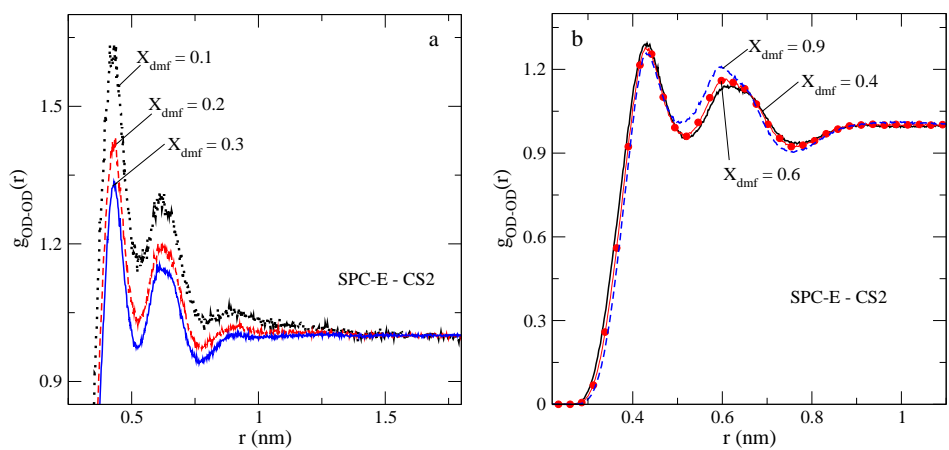

\begin{center}
\includegraphics[width=6cm,clip]{fig6a.eps}\quad
\includegraphics[width=6cm,clip]{fig6b.eps}
\end{center}
\caption{\label{fig4}(Color online) Evolution of the pair distribution functions,  OD-OD,
with changing solvent composition.}
\end{figure}

\begin{figure}[!b]
\begin{center}
\includegraphics[width=5.9cm,clip]{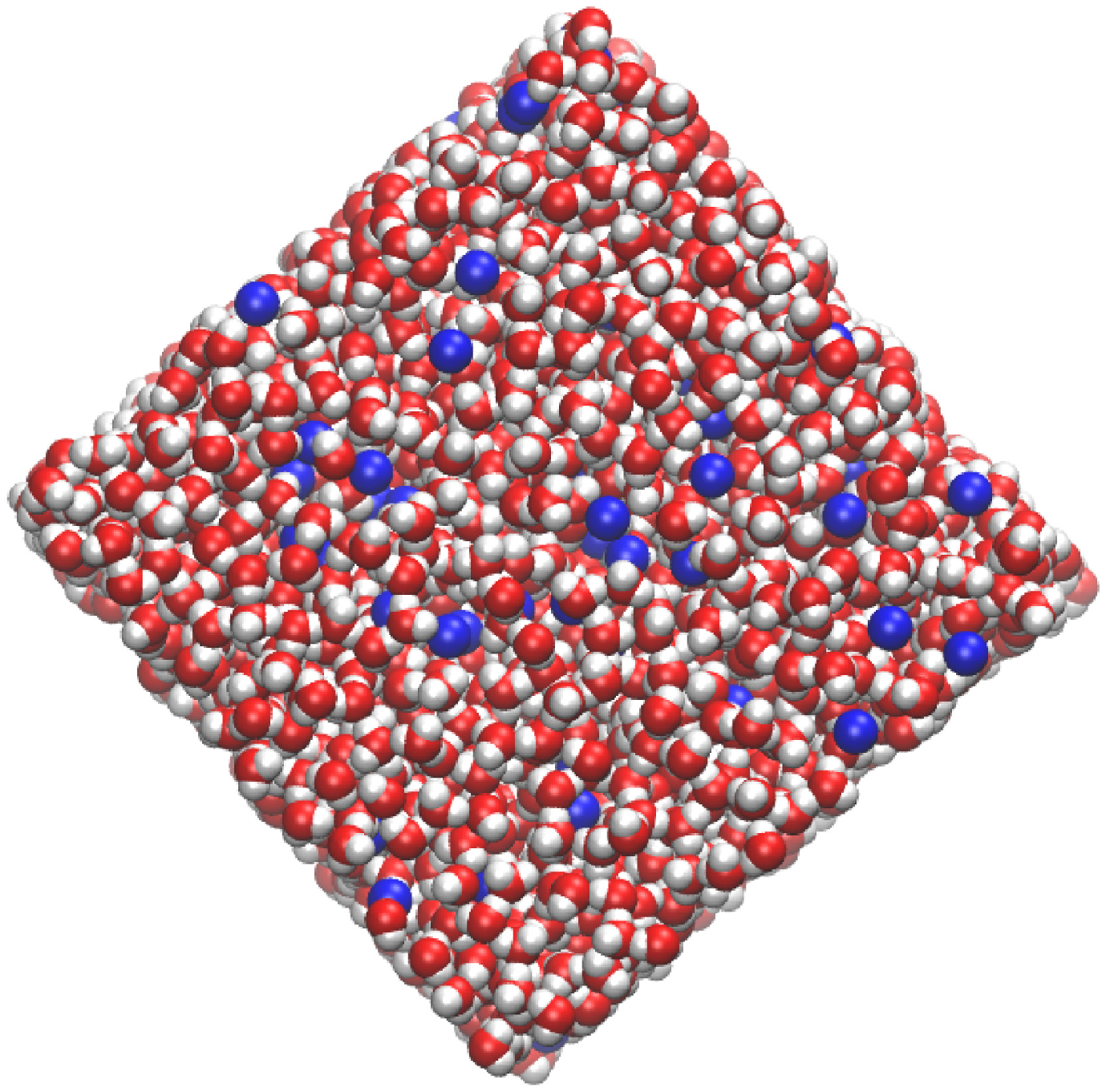}\quad
\includegraphics[width=5.4cm,clip]{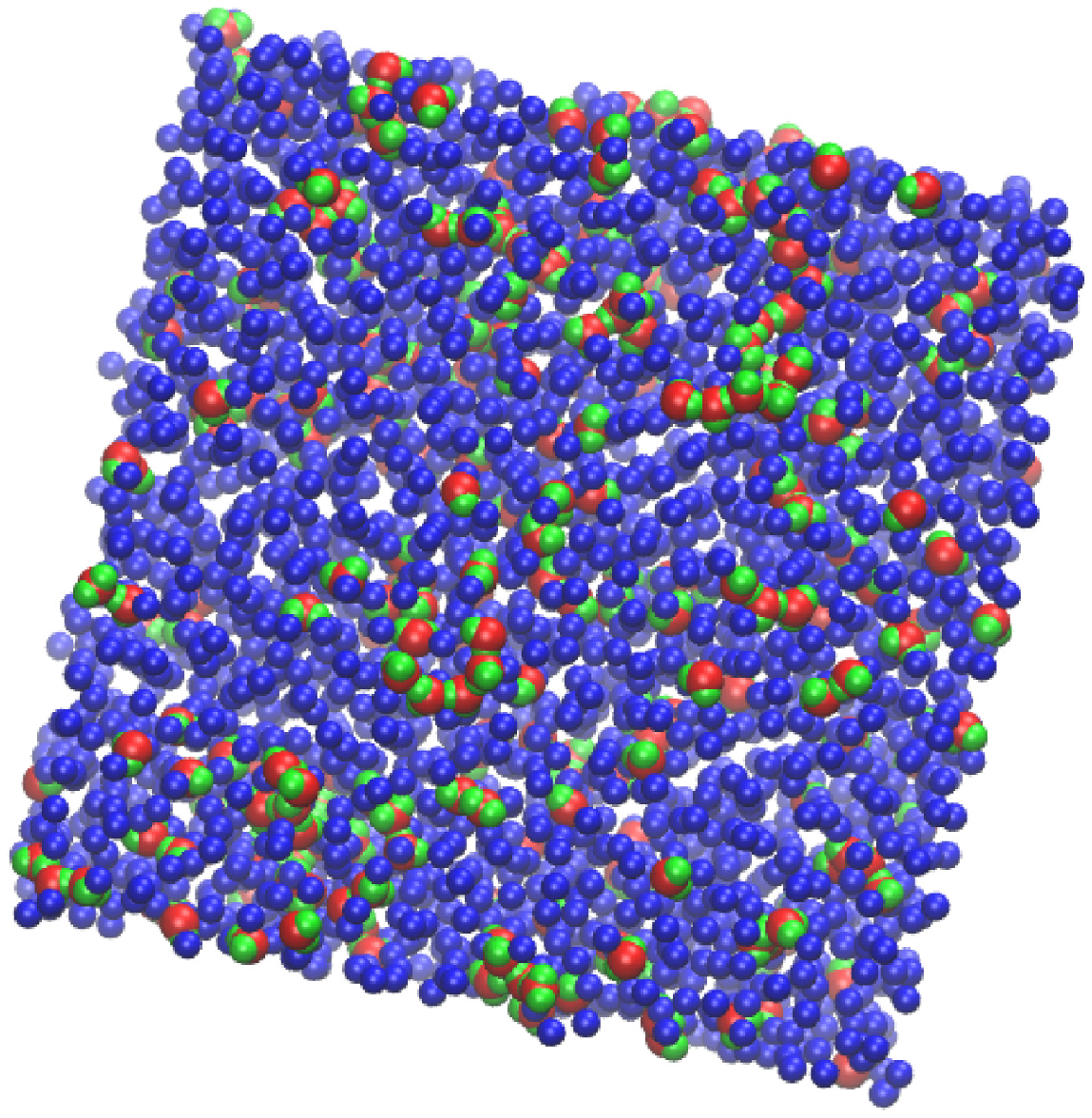}
\end{center}
\caption{\label{fig7}(Color online) Visualization of the distribution of DMF molecules (nitrogen atoms --- blue 
spheres) in water medium ($X_{\text{dmf}} = 0.03333$) in the left-hand panel (water oxygens and hydrogens
--- red and grey spheres, respectively), 
and of the distribution of water molecules in the DMF medium (nitrogens --- blue spheres)
at $X_{\text{dmf}} = 0.9$ in the right-hand panel 
(water oxygens and hydrogens --- red and green spheres, respectively).}
\protect
\end{figure}

The system is characterized by a large set of the site-site distribution functions. Our focus is
only in some of them, namely in the functions that describe most essential changes of the structure
of the system upon changes of its chemical composition. Evolution of the distribution of
DMF molecules on changes of $X_{\text{dmf}}$ is shown in terms of the distribution functions between DMF
oxygens, OD-OD (notation OD is used as an abbreviation for O$_{\text{dmf}}$), see 
two panels of figure~\ref{fig4}. If the DMF molecules are at small amount in the medium with
predominant number of waters,  the DMF particles seem preferring  to stay rather close
to one another, as it follows from the first two maxima of the pair distribution function (pdf) in panel~(a)
of figure~\ref{fig4}. However, even the first maximum is not high, the second one is much lower,
hence mutual contacts of two and more DMF molecules are not very probable.
If the concentration of DMF species increases, their distribution becomes even more
``homogeneous'', see, e.g., the curves are $X_{\text{dmf}} = 0.2$ and $0.3$ in panel~(a). In other words,
the first and second maxima as well as the minimum between them, all slightly decrease in magnitude and
at larger inter-particle separations, the pdf is close to unity.
This rather uniform distribution of DMF species, does not change significantly in an ample
interval of compositions, say from $X_{\text{dmf}} = 0.3$ up to $X_{\text{dmf}} = 0.9$. Moreover, we do not
observe a shift of the position of the first maximum of the pdf. It is located at $r \approx 0.43$~nm.
Only, the second maximum of the
OD-OD distribution slightly grows and shifts to smaller inter-particle separations when
the composition changes from $X_{\text{dmf}} = 0.4$ up to $X_{\text{dmf}} = 0.9$, indicating a weakly enhanced ``crowding''
of the DMF particles if their amount increases.
At this point we would like
to refer to the visualization of the  distribution of a small amount of DMF molecules
in the ``sea'' of waters, see figure~\ref{fig7}, panel~(a). This snapshot corresponds to the final configuration
of particles after $\approx 60$~ns.  
Both, the pdf at a low value of $X_{\text{dmf}}$ and the snapshot at similar conditions, permit to
conclude that there is no clustering of DMF species. 

\begin{figure}[!t]
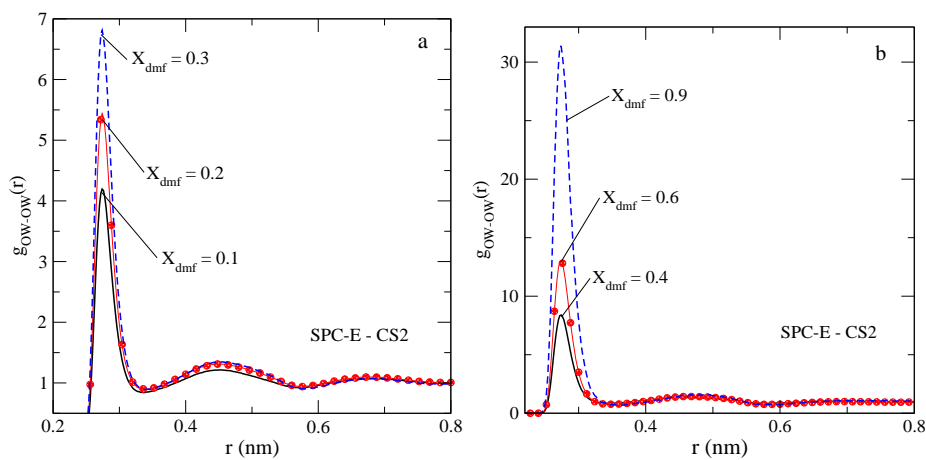

\begin{center}
\includegraphics[width=6cm,clip]{fig7a.eps}
\includegraphics[width=6cm,clip]{fig7b.eps}
\end{center}
\vspace{-2mm}
\caption{\label{fig5}(Color online) Evolution of the pair distribution functions,  OW-OW,
with changing solvent composition.}
\protect
\end{figure}

Let us proceed now to the description of the evolution of distribution of water molecules
in terms of the pdf OW-OW, figure~\ref{fig5}. The most pronounced trend is that the first maximum of 
$g_{\text{OW-OW}}$ strongly increases in magnitude upon increasing $X_{\text{dmf}}$, i.e., when the
concentration of water species decreases. However, important observation is that such a behavior is
not accompanied by a substantial growth of the second maximum of this function. To summarize,
the contacts between water molecules become more and more probable whereas the probabilistic
aspects of the structure of organic subsystem are less affected by changes of $X_{\text{dmf}}$. Water molecules
efficiently fill the space between the DMF particles, cf. the OW-OW first maximum grows at
$r \approx 0.28$~nm whereas the OD-OD first maximum remains almost intact at $r \approx 0.43$~nm.
The shape of OW-OW pdf together with the typical configuration of molecules given 
in the snapshot, right-hand panel of figure~\ref{fig7}, leads to the conclusion that water molecules are
rather uniformly distributed in the ``sea'' of DMF particles, closely situated waters 
are restricted to pairs at most.  

\begin{figure}[!t]
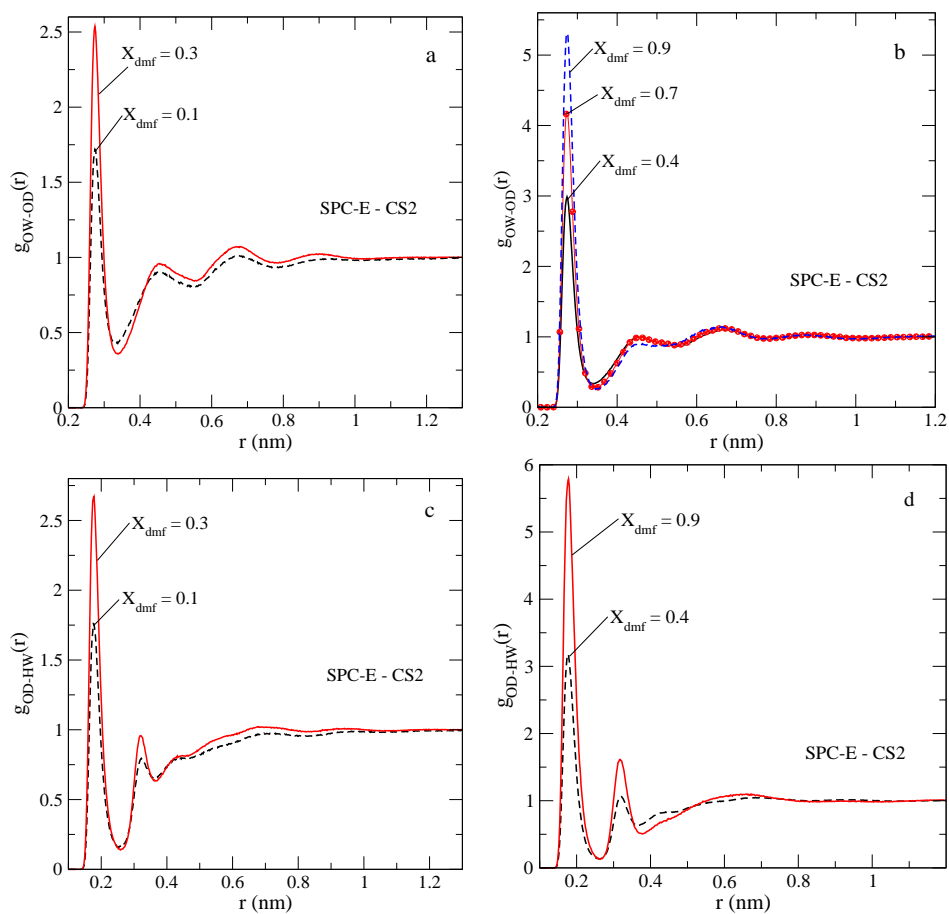

\begin{center}
\includegraphics[width=6cm,clip]{fig8a.eps}\quad
\includegraphics[width=6cm,clip]{fig8b.eps}
\\
\includegraphics[width=6cm,clip]{fig8c.eps}\quad
\includegraphics[width=6cm,clip]{fig8d.eps}
\end{center}
\vspace{-2mm}
\caption{\label{fig6} (Color online) Evolution of the pair distribution functions,  OW-OD,
and OD-HW with changing solvent composition.}
\protect
\end{figure}

Next, we would like to explore the trends of cross correlations between particles belonging to
different species in this binary mixture. We analyze them by using the OW-OD and OD-HW
distribution functions. A few representative examples are shown in figure~\ref{fig6}. The shape of the pdf
OW-OD changes principally at small interparticle separations. Namely, the first maximum 
increases in magnitude with an increasing $X_{\text{dmf}}$ in the entire range of composition, starting
from low values of $X_{\text{dmf}}$ up to the high values. Its position on the $r$-axis does not change, however.
As we have seen from figure~\ref{fig4}, the shape of OD-OD is weakly affected by $X_{\text{dmf}}$ values, thus changes
of the function OW-OD are quite small in the interval from $r \approx 0.4$~nm 
up to $r \approx 1$~nm.  In contrast to this behavior, the function OD-HW is sensitive to the value of
$X_{\text{dmf}}$ in the region of the first and of the second maximum. The first maximum of the OD-HW pdf
substantially increases in magnitude with an increasing $X_{\text{dmf}}$. Concerning the second maximum of this
function, it can be seen that its growth can be related to changes of the OW-OW pdf. Namely,
the second maximum of OD-HW becomes higher than unity for $X_{\text{dmf}} > 0.4$. Development of this
maximum can be attributed to an essential growth of OW-OW first maximum at a high concentration of
DMF molecules in the mixture. It is intuitively expected that hydrogen bonded  water molecules
occupying the space close to OD atom predominantly contribute to this second maximum of OD-HW,
because if hydrogens belonging to two different water molecules are close to OD, they should
contribute to the first maximum. 

\begin{figure}[!t]
\begin{center}
\includegraphics[width=6.5cm,clip]{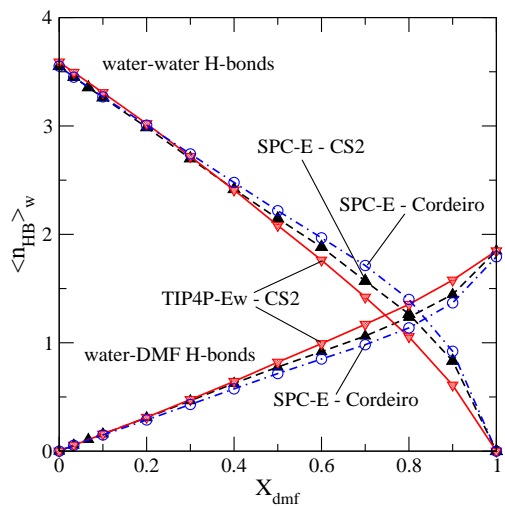}
\end{center}
\caption{\label{fig8} (Color online) Changes of the average number of water-water and water-DMF
hydrogen bonds (per water molecule) upon composition changes. The cross water-DMF
bonds for SPC-E-CS2 model are given by solid black triangles up.}
\protect
\end{figure}

Our final concern in this subsection is in the trends of behavior of the average number of hydrogen
bonds (normalization is performed per water molecule) between water molecules as well as
cross water-DMF bonds (figure~\ref{fig8}).  The GROMACS software using a default distance-angle
criterion was applied to count the numbers of bonded molecules in each case.
The behavior of $\langle n_{\text{HB}}\rangle$ as a function of $X_{\text{dmf}}$ is qualitatively similar to
what was recently discussed for other water-organic liquid mixtures~\cite{gujt1,gujt2}. 
The average number of H-bonds between water molecules decreases whereas the fraction of
the bonds between water molecules and DMF oxygen steadily increases with an increasing fraction of
DMF species. 
Only at very low values of $X_{\text{dmf}}$, the value for water-water $\langle n_{\text{HB}}\rangle$ is high,
yielding  evidence of the hydrogen bonded network between water molecules.
If DMF species are introduced into water, the number of cross contacts increases,
as it follows from the OW-OD and OD-HW functions, cf. panels~(a) and (d) of figure~\ref{fig6}. 
At high values of $X_{\text{dmf}}$, the average value of cross bonds becomes higher than
the one between waters. This is explicable straightforwardly, because at high concentration of DMF
species in the mixture there is no room to form fragments of expanded network of hydrogen bonds 
and, besides, the majority of water molecules are surrounded by DMF, either as individual entities or
very small groups such as pairs. It is difficult to establish precisely at what values of $X_{\text{dmf}}$
the hydrogen bonded network cease to exist as such. That would require additional and a more 
profound analysis of a set of topologic elements characterizing the cooperativity of bonds.
All three models explored in the present work (SPC-E-CS2, TIP4P-Ew-CS2 and SPC-E-Cordeiro), 
provide a qualitatively similar picture concerning $\langle n_{\text{HB}}\rangle(X_{\text{dmf}})$. Fine details depend
on water modelling and on the internal structure of DMF molecule. A model that predicts
slightly more probable water-water bonding in consequence leads to a slightly smaller fraction
of water-DMF bonds. For the moment, we have not explored the life-time and strength of the bonds 
formed in the system, however. 

Nevertheless, one more remark is pertinent. As we have seen  above, maxima on the 
experimentally measured excess mixing properties are observed in the 
interval $0.3 < X_{\text{dmf}} < 0.4$ (cf. figures~\ref{fig2} and \ref{fig3}) whereas the simulation predictions for the extrema
are in the interval $0.4 < X_{\text{dmf}} < 0.5$. This interval of compositions seems to correspond 
to the systems in which the effect of  hydrogen bonded network gradually ceases. It is tempting
to attribute this change of the behavior to the range of compositions in which the curves for 
$\langle n_{\text{HB}}\rangle(X_{\text{dmf}})$ start to deviate from linearity. However, this hypothesis should be explored and
possibly confirmed by  more sophisticated tools. Now, we would like to see if these observations of 
the trends of thermodynamic and structural properties are related to dynamic properties.

\subsection{Self-diffusion coefficients}

\begin{figure}[!b]
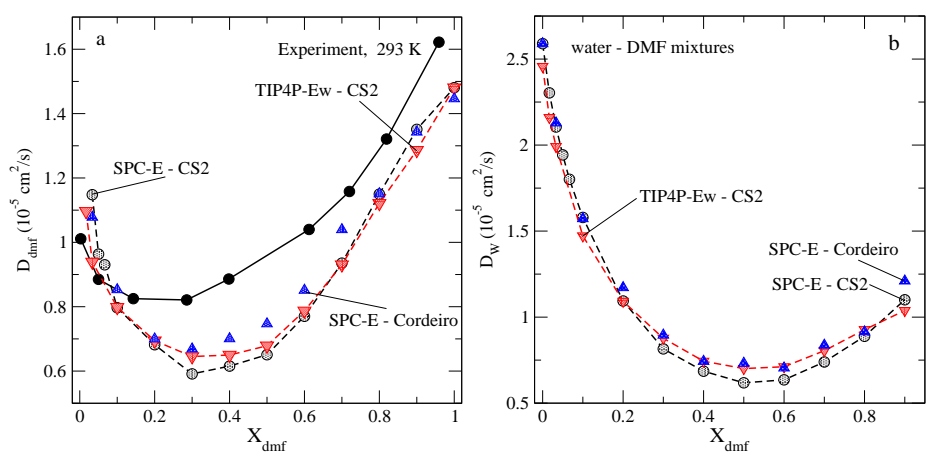

\begin{center}
\includegraphics[width=6cm,clip]{fig4a.eps}
\includegraphics[width=6cm,clip]{fig4b.eps}
\end{center}
\caption{\label{fig9} (Color online) Composition dependence of the self-diffusion coefficients of
species in water-DMF mixtures
from constant pressure-constant temperature simulations ($T = 298.15$~K, $P = 1$~bar).
In panel a the experimental data are taken from~\cite{volpe} at $T = 293$~K.}
\protect
\end{figure}

One of the principal objectives of the present study is to combine a discussion of an ample set of 
properties of water-DMF mixtures. Their dynamic properties undoubtfully deserve a separate work.
Here, we focus solely on the self-diffusion coefficients of water and DMF species. 
As usual, they are calculated from the mean-square displacement (MSD) of a particle via Einstein relation,
\begin{equation}
D_i =\frac{1}{6} \lim_{t \rightarrow \infty} \frac{\rd}{\rd t} \langle\vert {\bf 
r}_i(\tau+t)-{\bf r}_i(\tau)\vert ^2\rangle,
\end{equation}
where $i$ refers to water or DMF molecules and $\tau$ is the time origin.
Default settings of GROMACS were used for the separation of the time origins.
A set on trajectories coming from several consecutive simulations of $10$~ns were
combined to get an entire trajectory not less than 60--70~ns.
The fitting interval then has been chosen from $\approx10$\% to $\approx 50$\%
of the entire trajectory, to obtain  $D_{\text{dmf}}$ and $D_{\text w}$.
A set of results coming from the application of three models is given in two panels of figure~\ref{fig9}.
In spite of an extensive search in the literature, we have found only one set of experimental data
describing $D_{\text{dmf}}$ self-diffusion coefficient at $T = 293$~K. Our calculations refer to
$T = 298.15$~K, however. An overall shape of $D_{\text{dmf}}(X_{\text{dmf}})$ is similar from experiments and
from simulations of the models, panel~(a) of figure~\ref{fig9}. 
At each end of the curves there is a maximum value separated by a
minimum, the minimum in experiment is at $X_{\text{dmf}} \approx 0.2$ whereas all the simulated models
predict a minimum at $\approx 0.3$. Moreover, the models used in simulations yield lower values for
$D_{\text{dmf}}$ in a wide interval of compositions, from 0.1 to 1, comparing to experimental data. Only
at a rather low $X_{\text{dmf}}$, less than 0.1, the simulations predict higher values for $D_{\text{dmf}}$ than
from experiment. 

The self-diffusion coefficient of water species in binary water-DMF mixtures is presented in
panel~(b) of figure~\ref{fig9}. Unfortunately, we were unable to find the experimental data in spite of the efforts.
One well documented point is for pure water~\cite{vega}. Starting from this point, the values
for $D_{\text w}$ substantially decrease with an increasing $X_{\text{dmf}}$ till the minimum at $X_{\text{dmf}} = 0.5$. The
value of $D_{\text w}$ at minimum is approximately five times lower than for pure water.  
While $X_{\text{dmf}}$ increases further, the self-diffusion coefficient, $D_{\text w}$, increases in magnitude to
reach the values above unity (twice higher value than at minimum) for mixtures predominantly composed of
DMF species. Thus, for water-rich mixtures, $D_{\text w}$ is much higher than $D_{\text{dmf}}$, in contrast to
DMF-rich mixtures, where $D_{\text{dmf}}$ is higher than $D_{\text w}$. Note that the $y$-scale is different in two panels.
The magnitudes of $D_{\text{dmf}}$ and of $D_{\text w}$ at respective minima are close to each other. 

Huge difference between the self-diffusion coefficients of each species at two opposite sides of the
composition axis leads to the conclusion that there exists only a weak coupling of the dynamics of water
and DMF particles. At low $X_{\text{dmf}}$, the self-diffusion coefficient of water $D_{\text w}$ is approximately twice
higher than $D_{\text{dmf}}$. One can suppose that a molecule of  DMF is trapped in a water medium. The existence of
the hydrogen bond network does not suppress the self-diffusion of water molecules but contributes to
diminish self-diffusion of an individual DMF molecule in such a surrounding. On the other hand, 
on the opposite side, for DMF-rich mixtures, the self-diffusion of a water molecule is 
suppressed much less comparing to DMF, cf. ratio of $D_{\text{dmf}}$ and of $D_{\text w}$ at high $X_{\text{dmf}}$.  
The strongest coupling of the dynamics of two species is observed at intermediate values of composition
($0.3 < X_{\text{dmf}} < 0.5$), i.e., in the interval where the deviations from ideality of thermodynamic 
characteristics are at maximum. At these composition values, it seems that the effects of a hydrogen 
bonding between particles become much less important since the network cease to exist as such, and just
the packing and energetic aspects of mixing for systems close to equimolarity determine the minima
of the self-diffusion coefficients. 
It would be of interest to extend the exploration of the dynamic properties
by calculating viscosity and characteristic relaxation times for mixtures of different composition.
A broader set of properties --- stronger conclusions should come out. 

\subsection{Composition dependence of the dielectric constant}

\begin{figure}[!t]
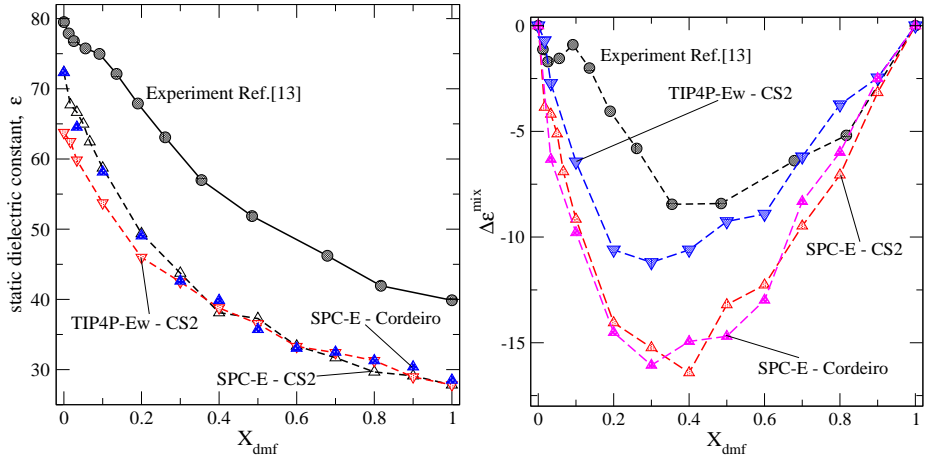

\begin{center}
\includegraphics[width=6cm,clip]{fig5a.eps}
\includegraphics[width=6cm,clip]{fig5b.eps}
\end{center}
\caption{\label{fig10} (Color online) Panel (a): Composition dependence of the static dielectric constant of
water-DMF mixtures
from constant pressure-constant temperature simulations ($T = 298.15$~K, $P = 1$~bar)
in comparison with the experimental data from~\cite{kumbharkhane}.
Panel~(b): Excess mixing static dielectric constant on composition in comparison with
experimental data.}
\protect
\end{figure}

Finally, we would like to put attention at one of various manifestations of the 
dielectric properties. The long-range, asymptotic behavior of correlations
between molecules possessing, e.g., a dipole moment is determined by the dielectric constant, $\varepsilon$.
It is commonly accepted that long molecular dynamics runs are necessary to obtain 
reasonable estimates for this property,
because $\varepsilon$ is calculated from the time-average of the fluctuations of the total
dipole moment of the system, 

\begin{equation}
\varepsilon=1+\frac{4\piup}{3k_{\text B}TV}\big(\langle\bf M^2\rangle-\langle\bf M\rangle^2\big),
\end{equation}
where $k_{\text B}$ is the Boltzmann constant and $V$ is the simulation cell volume. 
In this work we have taken 
care that each of the runs is of sufficient length, not less than $60$~ns in the majority of cases.
The curves coming out from simulations for three models of water-DMF  mixtures are shown in figure~\ref{fig10}~(a).
A general trend of the behaviour of $\varepsilon(X_{\text{dmf}})$ is that it decreases
with an increasing $X_{\text{dmf}}$, starting from a high value for pure water
to a lower value corresponding to pure DMF.
As it follows from the comparison of the simulation results and experimental data~\cite{kumbharkhane},
all three  models substantially underestimate the values for $\varepsilon$ in the entire
composition range. The static dielectric constant for pure DMF in the framework of CS2 model
is $\varepsilon \approx$ 27.83, whereas for the model of Cordeiro it is 28.52,
the experiment yields 39.88~\cite{kumbharkhane}. On the water-rich side, the discrepancy
between simulation data and experimental value evidently results from water modelling.
It is highly probable to improve the dependence of the static dielectric constant
on composition by applying the model for water specifically parametrized to reproduce
the dielectric constant~\cite{alejandre}. Simultaneously, it would require
parametrization of the force field for the DMF molecule. This task seems to be attainable.

Another sensitive test is provided by a comparison of the excess
dielectric constant [see equation~(\ref{eq1})], 
$\Delta \varepsilon^{\text{mix}} = \varepsilon_{\text m}- [X_{\text{dmf}}\varepsilon_{\text{dmf}}+(1-X_{\text{dmf}})\varepsilon_{\text w}]$,
with the experimental predictions~\cite{kumbharkhane}.
Experimental points indicate a negative deviation from
ideality in the entire composition range, figure~\ref{fig10}~(b). 
Interestingly, this behaviour is  contrary
to what follows for water-DMSO liquid mixtures~\cite{gujt1} but similar to what
we recently discussed for water-DME mixtures~\cite{gujt2}.
Maximal (negative) deviation from the ideal type behaviour
reported from the experimental measurements is at $X_{\text{dme}} \approx  0.45$.
The simulations results  reproduce the position of a minimum approximately; namely
the minimum is in the interval 0.3--0.4, dependent on the model employed.
These trends are in concordance with our observations concerning deviations from ideality
of thermodynamic properties and self-diffusion coefficients.
However, the magnitude of the excess static dielectric constant
is overestimated from the simulated models. The TIP4P-Ew combined with CS2 model is 
the most close to the experimental data. It is worth mentioning that the excess dielectric 
constant curve as a function of chemical composition of the mixture 
can be related to the excess refractive index measurements, see, e.g.,~\cite{gofurov}.
This issue has not been studied at the level of the united atom models for DMF so far.

All the above results have been obtained by using isobaric-isothermal simulations of
water-DMF mixtures. The final configurations of particles, however, can be conveniently used to
explore one of the interfacial properties by switching to the canonical, $NVT$, ensemble,
namely the surface tension on composition.

\subsection{Surface tension of water-DMF mixtures}

The simulations aiming in surface tension calculations at each point of composition 
axis have been performed by taking the final configuration of particles from the isobaric run.
Next, the box edge along $z$-axis was extended by a factor of 3, generating a box with
liquid slab and two liquid-mixture-vacuum interfaces in the $x$-$y$ plane, 
in close similarity to the procedure applied in \cite{fischer}. 
The total number of particles, three thousand, is reasonable
to yield an area of the $x$-$y$ face of the liquid slab sufficiently big. The elongation of
the liquid slab along $z$-axis is sufficient as well.
The executable molecular dynamics file was modified by
deleting a fixed pressure condition just preserving V-rescale thermostatting with the same
parameters as in $NPT$ runs. Other corrections have not been employed.

The values for the surface tension, $\gamma$, follow from the combination
of the time averages for the components of the pressure tensor,

\begin{equation}
\gamma =\frac {1}{2}L_z \Big\langle\Big[P_{zz} - \frac {1}{2} (P_{xx}+P_{yy})\Big]\Big\rangle,
\end{equation}  
where $P_{ij}$ ($i,j = x,y,z$) are the components of the pressure tensor, 
and $\langle\ldots\rangle$ denotes the time average.
We performed a set of runs at a constant volume, 5--6 each with the time duration of $10$~ns, 
and obtained the result for $\gamma$ by taking the block average. 

As in the previous subsections,
the results concern the SPC-E-CS2, SPC-E-Cordeiro and TIP4P-EW-CS2 models.
It is necessary to emphasize that our calculations were not focused to precisely reproduce
the values for $\gamma$ at a single point corresponding to pure water at a given density,
see, e.g.,~\cite{vega2}. We rather obtain $\gamma$ at a density of each mixture coming out
from the $NPT$ simulation. These values deviate from the experimental data as we 
have discussed in the corresponding subsection reporting the results. 
The experimental value of $\gamma$ reported in literature for water is 71.73~mN/m whereas
for DMF it is 34.4~mN/m. Our data for pure water models (SPC-E and TIP4P-Ew) agree
with the data given in~\cite{vega2} (see table~III and table~IV) prior the application
of the tail correction.

\begin{figure}[!t]
\begin{center}
\includegraphics[width=6.5cm,clip]{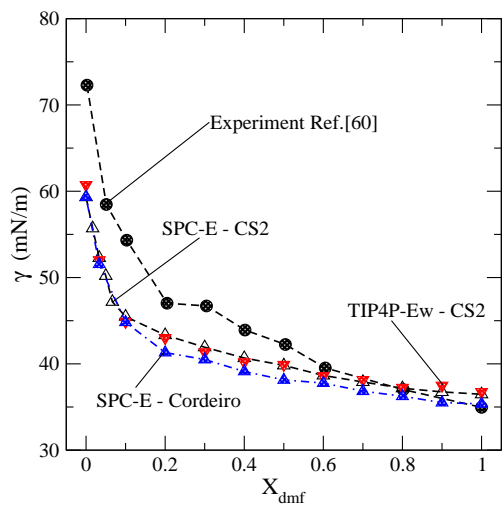}
\end{center}
\vspace{-2mm}
\caption{\label{fig11} (Color online) Composition dependence of the surface tension of
water-DMF mixtures
from constant volume-constant temperature simulations ($T = 298.15$~K)
in comparison with the experimental data from~\cite{forrest}.}
\protect
\end{figure}

A set of data obtained in the present simulation work is given in figure~\ref{fig11}, the experimental data 
are from the pioneering study~\cite{forrest}. The change of slope of the surface tension 
experimental curve
around 30~mole percent of DMF together with the maximum of the viscosity curve at this 
composition (not shown here), apparently may be related to the formation of a dihydrate of DMF.
Moreover, peculiarities of the mixing properties seen as a maximal deviation from
an ideal type of behavior and the minimum of $D_{\text{dmf}}$ support this assumption. We are not aware
of any spectroscopy results supporting this point of view.
 
However, general trends of the experimental dependence 
of the surface tension on $X_{\text{dmf}}$  are reproduced by the models used in simulations.
The values of $\gamma$ are underestimated in the water-rich mixtures. It seems that a better 
description of the surface tension of pure water should decrease this inaccuracy. By 
contrast, in the DMF-rich mixtures, the values for surface tension of mixtures are 
closer to the experimental data. The CS2 model for DMF satisfactorily describes the surface tension of a pure
organic liquid. The Cordeiro model also yields a reasonable value of the 
surface tension. On the other hand, the application of TIP4P-Ew water model does not
provide an improvement of the trends of behavior of $\gamma$, comparing to the SPC-E-CS2 model.   
Computer simulation results of this set of models do not evidence a peculiar behavior of
surface tension at $X_{\text{dmf}} \approx 0.3$.
We proceed now to the final comments, summary and conclusions.

\section{Summary and conclusions}

The mixtures explored in this work are an example of combined
water-organic liquid solvents. 
We performed extensive molecular dynamic simulations in
the isobaric-isothermal  ensemble to study the density, an ample set of mixing properties 
and the microscopic structure of  water-DMF mixtures in the entire range of
solvent composition. The self-diffusion coefficients of species and the
dielectric constant were calculated as well. 
All the simulations were performed at room temperature and ambient pressure,
1~bar.  Two water models (SPC-E and TIP4P-Ew), combined with the
united atom type CS2 and Cordeiro models for DMF
were studied. These insights are complemented by the surface tension simulations
for water-DMF mixtures using constant volume-constant temperature conditions.
The values for volume were chosen from the results of the isobaric simulations.

From a comparison with the available experimental data for different properties and
with the results of other authors on this and related systems,
we conclude that the predictions
obtained are qualitatively correct and give a physically sound picture of the properties explored.
All the properties investigated are sensitive to the
composition of a water-organic liquid solvent.

The principal conclusions of the present study can be resumed as follows.
We explored the evolution of the microscopic
structure in terms of the pair distribution functions. The simulation results
witness that the structure of the subsystem of DMF species is more
inert or less sensitive to the composition, in comparison to the structure
of an aqueous subsystem. 
The pair distribution functions evidence a
heterogeneous density distribution at local scale upon adding of DMF
molecules. These trends of changes of the microscopic structure bear similarity to
 water-DMSO and water-DME co-solvent mixtures, see, e.g.,~\cite{gujt1,gujt2}.
Possibly  the ``associated'' species involving water and DMF molecules 
can be formed in the system, with or without hydrogen bonds. This observation
is in accordance with the interpretation of experimental data.
Water-water and water-organic co-solvent average number of hydrogen bonds 
do not show a peculiar
behaviour, in comparison to qualitatively similar  mixtures of water with 
organic co-solvent~\cite{gujt1,gujt2}.

Dynamic properties are discussed
in terms of the self-diffusion coefficients of species, $D_{\text w}$ and $D_{\text{dmf}}$. Both
of them exhibit a minimum in the interval of composition that corresponds to
most ``packed'' structures according to the behaviour of a mixing volume. The values
of self-diffusion coefficients of the species for pure components are reasonably well
described. Concerning the dependence of the static dielectric constant on composition,
we observe that its excess is better described while using the TIP4P-Ew water model 
combined with the CS2 model for DMF. It would be of interest to relate the behaviour of
the static dielectric constant on the composition with refractive index and viscosity data.
Finally, we obtained a reasonable description of the dependence of surface tension on composition,
a better agreement with experimental data is expected from application of other water models. 

At the present stage of the development, there are several  missing elements 
worthwhile a detailed investigation. Namely, the application of  all-atom force 
fields for the DMF molecules for the description of water-DMF mixtures will be
reported in a separate publication. A comparison of united-atom type models
with all-atom modelling is of primordial importance. 
On the other hand, insights into the behaviour of
dynamic and dielectric properties by exploration of, e.g., the relaxation times, viscosity,
hydrogen-bonds life-time and complex dielectric constant,  would be desirable.
Further research is planned in exploration of solutions with 
complex molecules in water-DMF mixtures along the lines of the available in literature 
experimental research. 

\section*{Acknowledgements}

O.P. is grateful to D. Vazquez and M. Aguilar for technical support of this work
at the Institute of Chemistry of the UNAM.

\ukrainianpart
\title{Залежність від концентрації мікроскопічної структури, термодинамічних, динамічних та електричних властивостей модельної суміші
вода-диметил формамід. Результати симуляцій методом молекулярної динаміки}

\author{Г. Домінгес, O. Пізіо}
\address{Iнститут матерiалознавства, Нацiональний автономний унiверситет м. Мехiко, Мехiко, Мексика
}

\makeukrtitle

\begin{abstract}
Здійснено симуляції методом молекулярної динаміки в ізотермічно-ізобаричному ансамблі з метою вивчення широкого набору властивостей моделі суміші
вода-N,N-диметилформамід (DMF)  як функцій концентрації. Використано моделі води SPC-E і TIP4P-Ew разом з двома об'єднаними атомними моделями для DMF
[Chalaris M., Samios J., J. Chem. Phys., 2000, \textbf{112}, 8581;
Cordeiro J., Int. J. Quantum Chem., 1997, \textbf{65}, 709].
Наш основний  аналіз стосується поведінки структурних властивостей в термінах радіальних функцій розподілу і числа водневих зв'язків між
молекулами різних сортів, а також термодинамічних властивостей. Зокрема, ми досліджуємо густину, надлишкові молярний об'єм та ентальпію змішування, питому теплоємність і надлишкову питому теплоємність змішування. Накінець, обговорюються коефіцієнти самодифузії сортів і діелектрична стала системи. Крім того, обчислюється та аналізується поверхневий натяг
 сумішей вода-DMF.

\keywords моделі води, моделі N,N-диметилформаміду, термодинамічні властивості, коефіцієнт самодифузії, діелектрична стала, поверхневий натяг, молекулярна динаміка
\end{abstract}


\begin{thebibliography}{99}

\bibitem{kroschwitz} Kroschwitz J.I., Seidel A. (Eds.),
Kirk-Othmer Encyclopedia of Chemical Technology, Vol.~1, Wiley-Interscience, Hoboken, 2004.

\bibitem{fiorito} Fiorito A., Larese F., Molinari S., Zanin T., Am. J. Ind. Med., 1997,
\textbf{32}, 255, \\
      \doi{10.1002/(SICI)1097-0274(199709)32:3<255::AID-AJIM11>3.0.CO;2-U}.

\bibitem{ohara} Ohara M., Takagaki A., Nishimura S., Ebitani K., Appl. Catal. A, 2010,
\textbf{383}, 149, \\
      \doi{10.1016/j.apcata.2010.05.040}.

\bibitem{ohtaki} Ohtaki H., Itoh S., Yamaguchi T., Ishiguro S., Rode B.M.,
Bull. Chem. Soc. Jpn., 1983, \textbf{56}, 3406, \\
      \doi{10.1246/bcsj.56.3406}.

\bibitem{okada} Okada M., Ibuki K., Ueno M., Bull. Chem. Soc. Jpn., 2012, \textbf{85}, 189,
      \doi{10.1246/bcsj.20110233}.

\bibitem{biliskov} Bili\v{s}kov N., Baranovi\'c G., J. Mol. Liq., 2009, \textbf{144}, 155,
      \doi{10.1016/j.molliq.2008.11.004}.

\bibitem{schoester} Schoester P.C., Zeidler M.D., Radnai T., Bopp P.A., Z. Naturforsch. A: Phys. Sci., 1995,
\textbf{50}, 38,\\
      \doi{10.1515/zna-1995-0106}.

\bibitem{guarino} Guarino G., Ortona O., Sartorio R., Vitagliano V.,
J. Chem. Eng. Data, 1985, \textbf{30}, 366,
      \doi{10.1021/je00041a039}.

\bibitem{volpe} Volpe C.D., Guarino G., Sartorio R., Vitagliano V.,
J. Chem. Eng. Data, 1986, \textbf{31}, 37,
      \doi{10.1021/je00043a012}.

\bibitem{bernal} Bernal-Garc\'ia J.M., Guzm\'an-L\'opez A., Cabrales-Torres A.,
Estrada-Baltazar A., Iglesias-Silva G.A., J. Chem. Eng. Data, 2008,
\textbf{53}, 1024,
      \doi{10.1021/je700671t}.

\bibitem{visser} De Visser C., Perron G., Desnoyers J.E.,
Heuvelsland W.J.M., Somsen G., J. Chem. Eng. Data, 1977, \textbf{22}, 74,
      \bibdoi{10.1021/je60072a016}.

\bibitem{cilense} Cilense M., Benedetti A.V., Vollet D.R., Thermochim. Acta, 1983, \textbf{63}, 151,\\
      \bibdoi{10.1016/0040-6031(83)80080-X}.

\bibitem{kumbharkhane} Kumbharkhane A.C., Puranik S.M., Mehrotra S.C., J. Solution Chem., 1993,
\textbf{22}, 219,
      \bibdoi{10.1007/BF00649245}.

\bibitem{checoni} Checoni R.F., Volpe P.L.O., J. Solution Chem., 2010, \textbf{39}, 259,
      \bibdoi{10.1007/s10953-010-9500-6}.

\bibitem{gofurov} Gofurov Sh., Ismailova O., Makhmanov U., Kokhkharov A.,
Int. J. Chem. Mol. Nucl. Mater. Metall. Eng., 2017, \textbf{11}, 330.

\bibitem{ueno} Ueno M., Mitsui R., Iwahashi H., Tsuchihashi N., Ibuki K.,
J. Phys. Conf. Ser., 2010, \textbf{215}, 1, \\
      \bibdoi{10.1088/1742-6596/215/1/012074}.

\bibitem{scharlin} Scharlin P., Steinby K., Doma\'nska U., J. Chem. Thermodyn., 2002,
\textbf{34}, 927,
      \bibdoi{10.1006/jcht.2002.0946}.

\bibitem{han} Han K.-J., Oh J.-H., Park S.-J., Gmehling J.,
J. Chem. Eng. Data, 2005, \textbf{50}, 1951,
      \bibdoi{10.1021/je050209y}.

\bibitem{garcia} Garc\'ia B., Alcalde R., Leal J.M., Matos J.S., J. Phys. Chem. B, 1997,
\textbf{101}, 7991,
      \bibdoi{10.1021/jp9626374}.

\bibitem{shokouhi} Shokouhi M., Jalili A.H., Hosseini-Jenab M., Vahidi M.,
J. Mol. Liq., 2013, \textbf{186}, 142,\\
      \bibdoi{10.1016/j.molliq.2013.07.005}.

\bibitem{chen} Chen L., Gro{\ss} T., L\"udemann H.-D., Z. Phys. Chem., 2000, \textbf{214}, 239,
      \bibdoi{10.1524/zpch.2000.214.2.239}.

\bibitem{jadzyn} Jad\.zyn J., \'Swiergiel J., Phys. Chem. Chem. Phys., 2012, \textbf{14}, 3170,
      \bibdoi{10.1039/C2CP23960D}.

\bibitem{egorov} Egorov G.I., Makarov D.M., Kolker A.M., Russ. J. Phys. Chem. A, 2007,
\textbf{81}, 528,\\
      \bibdoi{10.1134/S003602440704005X}.

\bibitem{bai1} Bai T.-C., Yao J., Han S.-J., J. Chem. Thermodyn., 1998, \textbf{30}, 1347,
      \bibdoi{10.1006/jcht.1998.0402}.

\bibitem{bai2} Bai T.-C., Yao J., Han S.-J., Fluid Phase Equilib., 1998, \textbf{152}, 283,
      \bibdoi{10.1016/S0378-3812(98)00402-6}.

\bibitem{schmid1} Schmid E.D., Brodbek E., J. Chem. Phys., 1983, \textbf{78}, 1117,
      \bibdoi{10.1063/1.444895}.

\bibitem{schmid2} Schmid E.D., Brodbek E., J. Mol. Struct. THEOCHEM, 1984, \textbf{108}, 17,
      \bibdoi{10.1016/0166-1280(84)80095-0}.

\bibitem{jorgensen} Jorgensen W.L., Swenson C.J., J. Am. Chem. Soc., 1985, \textbf{107}, 569,
      \bibdoi{10.1021/ja00289a008}.

\bibitem{yashonath} Yashonath S., Rao C.N.R., Chem. Phys., 1991, \textbf{155}, 351,
      \bibdoi{10.1016/0301-0104(91)80111-T}.

\bibitem{cordeiro1} Cordeiro J.M.M., Int. J. Quantum Chem., 1997, \textbf{65}, 709,\\
      \bibdoi{10.1002/(SICI)1097-461X(1997)65:5<709::AID-QUA37>3.0.CO;2-U}.

\bibitem{cordeiro2} Cordeiro J.M.M., Freitas L.C.G., Z. Naturforsch. A: Phys. Sci., 1999,
\textbf{54}, 110,
      \bibdoi{10.1515/zna-1999-0204}.

\bibitem{cordeiro3} Cordeiro M.A.M., Santana W.P., Cusinato R., Cordeiro J.M.M.,
J. Mol. Struct. THEOCHEM, 2006, \textbf{759}, 159,
      \bibdoi{10.1016/j.theochem.2005.11.016}.

\bibitem{bako} Radnai T., Bak\'o I., Jedlovszky P., P\'alink\'as G., Mol. Simul., 1996, \textbf{16}, 345,
      \bibdoi{10.1080/08927029608024084}.

\bibitem{chalaris1} Chalaris M., Samios J., J. Chem. Phys., 2000, \textbf{112}, 8581, 
      \bibdoi{10.1063/1.481460}.

\bibitem{chalaris2} Chalaris M., Samios J., J. Mol. Liq., 1998, \textbf{78}, 201,
      \bibdoi{10.1016/S0167-7322(98)00092-0}.

\bibitem{chalaris3} Chalaris M., Koufou A., Samios J., J. Mol. Liq., 2002, \textbf{101}, 69,
      \bibdoi{10.1016/S0167-7322(02)00103-4}.

\bibitem{zoranic1} Zorani\'c L., Mazighi R., Sokoli\'c F.,  Perera A., J. Phys. Chem. C, 2007, 
\textbf{111}, 15586,
      \bibdoi{10.1021/jp0736894}.

\bibitem{zoranic2} Zorani\'c L., Mazighi R., Sokoli\'c F.,  Perera A., J. Chem. Phys., 2009,
\textbf{130}, 124315,
      \bibdoi{10.1063/1.3093071}.

\bibitem{gao} Gao J., Pavelites J.J., Habibollazadeh D., J. Phys. Chem., 1996,
\textbf{100}, 2689,
      \bibdoi{10.1021/jp9521969}.

\bibitem{biswas} Biswas S., Mallik B.S., J. Chem. Eng. Data, 2014, \textbf{59}, 3250,
      \bibdoi{10.1021/je5002544}.

\bibitem{vasudevan} Vasudevan V., Mushrif S.H., J. Mol. Liq., 2015, \textbf{206}, 338,
      \bibdoi{10.1016/j.molliq.2015.03.004}.

\bibitem{lei} Lei Y., Li H., Pan H., Han S., J. Phys. Chem. A, 2003, \textbf{107}, 1574,
      \bibdoi{10.1021/jp026638+}.

\bibitem{jia} Jia G.-Z., Huang K.-M., Yang L.-J., Yang X.-Q., Int. J. Mol. Sci., 2009, \textbf{10}, 1590,
      \bibdoi{10.3390/ijms10041590}.

\bibitem{razzokov} Razzokov D., Ismailova O.B., Mamatkulov Sh.I., Trunilina O.V., Kokhkharov A.M.,
Russ. J. Phys. Chem. A, 2014, \textbf{88}, 1500,
      \bibdoi{10.1134/S0036024414090271}.

\bibitem{fischer} Fischer N.M., van Maaren P.J., Ditz J.C., Yildirim A.,
van der Spoel D., J. Chem. Theory Comput., 2015, \textbf{11}, 2938,
      \bibdoi{10.1021/acs.jctc.5b00190}.

\bibitem{galicia1} Galicia-Andr\'es E., Pusztai L., Temleitner L., Pizio O.,
J. Mol. Liq., 2015, \textbf{209}, 586,\\
      \bibdoi{10.1016/j.molliq.2015.06.045}.

\bibitem{galicia2} Galicia-Andr\'es E., Dominguez H., Pusztai L., Pizio O., Condens. Matter Phys., 2015,
\textbf{18}, 43602,
      \bibdoi{10.5488/CMP.18.43602}.

\bibitem{gujt1} Gujt J., C\'azares Vargas E., Pusztai L., Pizio O., J. Mol. Liq., 2017, \textbf{228}, 71,
      \bibdoi{10.1016/j.molliq.2016.09.024}.

\bibitem{gujt2} Gujt J., Dominguez H., Sokolowski S., Pizio O., Condens. Matter Phys., 2017,
\textbf{20},  33603,\\
      \bibdoi{10.5488/CMP.20.33603}.
\bibitem{takamuku1} Takamuku T., Shimomura T., Tachikawa M., Kanzaki R., Phys. Chem. Chem. Phys., 2011,
\textbf{13}, 11222,
      \bibdoi{10.1039/C0CP00338G}.

\bibitem{takamuku2} Takamuku T., Yamaguchi A., Matsuo D., Tabata M., Kumamoto M., Nishimoto J.,
Yoshida K., Yamaguchi T., Nagao M., Otomo T., Adachi T., J. Phys. Chem. B, 2001, \textbf{105}, 6236,
     \bibdoi{10.1021/jp003011n}.

\bibitem{takamuku3} Takamuku T., Yamaguchi A., Matsuo D., Tabata M., Yamaguchi T.,
Otomo T., Adachi T., J. Phys. Chem. B, 2001, \textbf{105}, 10101,
      \bibdoi{10.1021/jp011692w}.

\bibitem{takamuku4} Takamuku T., Noguchi Y., Yoshikawa E., Kawaguchi T., Matsugami M.,
Otomo T., J. Mol. Liq., 2007, \textbf{131--132}, 131,
      \bibdoi{10.1016/j.molliq.2006.08.048}.

\bibitem{gromacs} Van der Spoel D., Lindahl E., Hess B., Groenhof G., Mark A.E.,
Berendsen H.C., J. Comput. Chem., 2005, \textbf{26}, 1701,
      \bibdoi{10.1002/jcc.20291}.

\bibitem{spce} Berendsen H.J.C., Grigera J.R., Straatsma T.P., J. Phys. Chem., 1987, \textbf{91}, 6269,
      \bibdoi{10.1021/j100308a038}.

\bibitem{horn} Horn H.W., Swope W.C., Pitera J.W., Madura J.D., Dick T.J., Hura G.L.,
Head-Gordon T., J. Chem. Phys., 2004, \textbf{120}, 9665,
      \bibdoi{10.1063/1.1683075}.

\bibitem{vega} Vega C., Abascal J.L.F., Phys. Chem. Chem. Phys., 2011, \textbf{13}, 19663,
      \bibdoi{10.1039/C1CP22168J}.

\bibitem{wensink} Wensink E.J.W., Hoffmann A.C., van Maaren P.J., van der Spoel D.,
J. Chem. Phys., 2003, \textbf{119}, 7308,
      \bibdoi{10.1063/1.1607918}.

\bibitem{alejandre} Alejandre J., Chapela G.A., Saint-Martin H., Mendoza H., Phys. Chem. Chem. Phys., 2011,
\textbf{13}, 19728,
      \bibdoi{10.1039/C1CP20858F}.

\bibitem{vega2} Vega C., de Miguel E., J. Chem. Phys., 2007, \textbf{126}, 154707,
      \bibdoi{10.1063/1.2715577}.

\bibitem{forrest} Blankenship F., Clampitt B., Proc. Oklahoma Acad. Sci., 1950, \textbf{31}, 106.

\end{thebibliography}
\end{document}